\documentclass{ws-ijmpd}

\begin{document}

\markboth{Ulisses Barres de Almeida.}
{H.E.S.S. Observations of Relativistic Sources}

%
\catchline{}{}{}{}{}
%

\title{RESULTS FROM H.E.S.S. OBSERVATIONS OF RELATIVISTIC SOURCES 
  }

\author{Ulisses Barres de Almeida}

\address{Department of Physics, University of Durham, South Road\\
Durham, DH1 3LE, England
\\u.b.almeida@dur.ac.uk}

\author{for the H.E.S.S. Collaboration}


\maketitle

\begin{history}
\received{Day Month Year}
\revised{Day Month Year}
\comby{Managing Editor}
\end{history}

\begin{abstract}
The High Energy Stereoscopic System (H.E.S.S.) is a southern
hemisphere array of four Imaging Atmospheric Cherenkov Telescopes
observing the sky in the very high energy gamma-ray range (E $>$
100 GeV). VHE observations are an invaluable tool to study the
acceleration and propagation of energetic particles in many
astrophysical systems where relativistic outflows are the main drivers
of the emission, such as AGNs and galactic binary systems. In
this paper the main results of the H.E.S.S. observations of these
objects will be reviewed, and the general picture that emerges from
them will be presented. We will also comment on prospects for
future investigations with H.E.S.S.-II.
\end{abstract}


\section{Introduction}	
This paper is a short review of what we consider to be the most relevant
results obtained by the H.E.S.S. collaboration from
relativistic sources in recent years. We report on
both galactic and extragalactic sources, concentrating on those
objects whose behaviour most add to or most
challenge our understanding of high-energy processes in astrophysics. 
Relativistic outflows are the environment of extreme physics \textit{par
  excellence}, being the locus of multi-TeV particle acceleration in a
number of distinct astrophysical systems. The rapidly developing field of
ground-based gamma-ray astronomy provides a window into the
very-high energy (VHE; $E >$ 100 GeV) universe\cite{ARAA}, furnishing
us with direct information on the populations of energetic particles in objects
such as Active Galactic Nuclei (AGN), pulsar winds and galactic binary
systems.

\section{The High Energy Stereoscopic System}

The H.E.S.S. array of four large ($\sim$ 100 m$^2$ collecting area)  
Imaging Atmospheric Cherenkov Telescopes (IACTs), located in the Khomas
highlands of Namibia ($23.3^{\circ}$ S,$15.5^{\circ}$ E; 1800 m
a.s.l.), has been observing the sky since
2004\cite{hess}. Ground-based gamma-ray instruments detect the 
Cherenkov light emitted by the secondary particles of $\gamma$-ray-initiated 
air showers. The array operates on the concept of a stereoscopic system of
IACTs, whereby multiple telescopes are used to simultaneously
image the air showers from different viewing angles to improve energy
resolution, reconstruction of the primary photon direction ($\sim 0.05^{\circ} -
0.1^{\circ}$) and rejection of the dominant cosmic-ray background ($
> 99\%$). The H.E.S.S. camera F.O.V. of $\sim 5^{\circ}$ is suited for
imaging the (usually) extended galactic sources and facilitates the execution of
extended surveys, such as that of the Galactic plane\cite{survey}. The
system is also capable of spectral measurements with an energy
resolution of about 15\%. With a post-background rejection
energy threshold of $\sim$ 120 GeV, and an effective area of about
10$^5$ m$^2$ (both at Zenith), point sources with fluxes of the
order of 10 mCrab ($3\times10^{-13}$ erg cm$^{-2}$ s$^{-1}$) are
detected within 50 h of observations.

\section{Extragalactic Relativistic Sources}

The rapidly populating TeV sky today numbers over 80 detected sources 
(see http://tevcat.uchicago.edu/), including a number
of different object classes. Greater typological diversity is found
amongst the Galactic population, where compact objects, binary systems
and massive stars at late evolutionary stages figure as the primary
sites of particle acceleration in the Galaxy. The extraglactic sources
seen by H.E.S.S. are almost exclusively AGNs (10 BL Lacs + 2 FRI radio
galaxies), the exception being the newly detected starburst galaxy NGC
253\cite{NGC253}, whose emission is believed to originate from
the combined activity of a large number of supernovae in its
central regions rather than from nuclear activity.

\subsection{PKS 2155-304 as viewed in TeVs}

The high-frequency peaked BL Lac (HBL) PKS 2155-304\cite{chadwick} (z
= 0.116) has been a primary target for H.E.S.S. since the start of its
operations, and four years of continual monitoring  make it one of the
best-studied extragalactic VHE source in the sky. Its
brightness at TeV energies ($\sim$ 0.15 Crab low flux state)
guarantees the source is detected whenever observed by H.E.S.S., 
allowing the derivation of a well-sampled
long-term light-curve which presents variability on timescales of days to 
years\cite{punch}. Here we report on the source's
extreme outburst observed by H.E.S.S. in July 2006 and on 
observations performed jointly with the Fermi satellite in 2008.        

\begin{figure}[pt]
\centerline{\epsfig{file=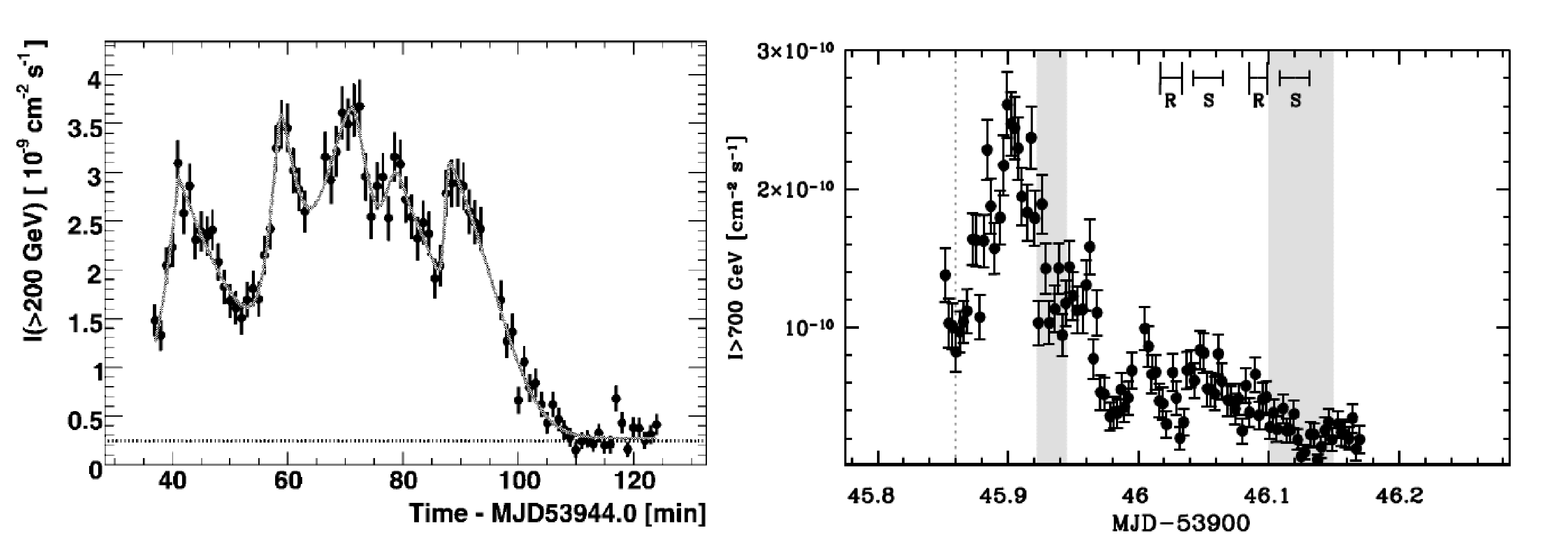, width=12.5cm}}
\vspace*{8pt}
\caption{Intranight VHE light curves of the two extreme flares of 
PKS 2155-304 on 28\protect\cite{largeflare} (left; $> 200$ GeV) and
30\protect\cite{chandranight} (right; $> 700$ GeV) July 2006.}
\label{fig1}
\end{figure}

\subsubsection{Extreme outburst events in July 2006}

The HBL PKS 2155-304 was detected by H.E.S.S. in an unusually high
state throughout the month of July 2006, with an average baseline flux
of about $2\times 10^{-10}$ cm$^{-2}$ s$^{-1}$, 5$\times$ higher than
the typical low state for the source. Its activity
peaked in two extreme flares observed on the nights of 28 and 30
July (MJD 53944 and 53946). The first flare\cite{largeflare} reached a
maximum flux ($>$ 200 GeV) of about 15.1 Crab, almost 2 orders of
magnitude above the low state, with a nightly average flux of $\sim
1.7\times 10^{-9}$  cm$^{-2}$ s$^{-1}$. Intranight variability episodes with
doubling timescales of $\sim$ 2 min, the fastest ever
seen for an AGN at any wavelength, were registered (see
Fig.~\ref{fig1}). The power spectrum distribution (PSD) showed
significant variability above the noise level up to $1.6\times10^{-3}$
Hz (600 s); the Fourier spectrum index of $-2.06\pm0.2$ indicates that
most of the power is present at low frequencies and that the
variability behaviour is compatible with a stochastic process. These
PSD characteristics are very similar to those derived
in X-rays for the same source, although the fractional rms variability 
amplitude is significantly higher in $\gamma$-rays\cite{zhang}.

Causality arguments bound the size $R$ of the emission zone to the
variability timescale $t_{\rm{var}}$ and the Doppler factor $\delta$ of the
flow. If the jets of blazars are indeed powered by accretion onto a
supermassive black hole, the Schwarzschild radius of the central object sets the
fundamental scale for the \textit{global} dynamics of the system. 
The shortest value $t_{\rm{var}} \sim 170$ s registered implies a size
for the emission region $R\delta^{-1} \le 4.65\times10^{12}$ cm $\le
0.31$ AU, so that for $R>R_{\rm{S}}$ ($ M_{\rm{SMBH}} ~ 1-2\times
10^{9} M_{\odot}$)\cite{bettoni} Doppler factors as
large as 50 are required, much higher than
typically derived for blazars or what can be easily accommodated by
theory. This unprecedented constraint on the location and dynamics of
particle acceleration sites in extragalactic jets suggests
that the VHE variability is not \textit{directly} tied to the 
central engine and that the VHE emission originates from a distinct, 
more compact and energetic population within the flow\cite{begelman}.

\begin{figure}[pt]
\centerline{\epsfig{file=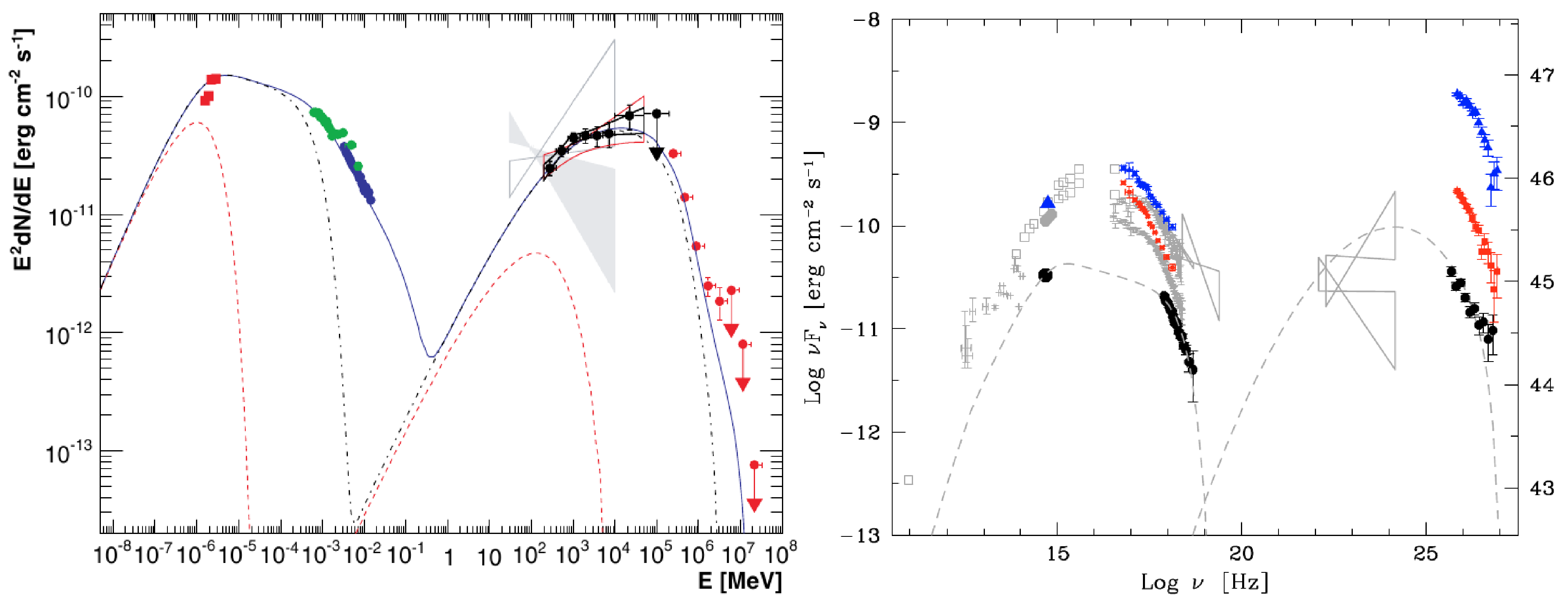, width=12.3cm}}
\vspace*{8pt}
\caption{Spectral Energy Distribution of PKS 2155-304. Left: Low-state
  SED during simultaneous H.E.S.S. and Fermi observations; the
  continuous curve is a single-zone SSC model fit to the SED 
  (with $\Gamma \sim$ 30 and $B \sim$ 0.02). The red-dashed and the
  dot-dashed lines are for the same model without electrons above
  $\gamma_{1} \sim 1\times 10^{4}$ and $\gamma_{2} \sim 2\times
  10^{5}$\protect\cite{hessfermi}.
  Right: Simultaneous X-ray/VHE high-state SED for the night of 30
  July 2006; notice the Compton dominance during the high state and
  the presence of no shifts on the overall SED peaks during
  variability\protect\cite{chandranight}.}
\label{fig2}
\end{figure}

On the second night of extreme variability, observations were
conducted as part of a MWL campaign with Chandra; the VHE flux
attained values comparable to those of the first
flare\cite{chandranight}. The intranight light-curve showed a large
VHE outburst at the start of observations, which then faded along the
rest of the night. The X-ray flux variability closely matched the VHE 
behaviour, whereas the simultaneous optical light-curve presented
less variability and a more persistent high flux level. A novel and
important feature of this night is the large Compton dominance of the
source's high state ($L_{C}/L_{S} \sim 8-10$), which is apparent in
the SED of Fig.~\ref{fig2}; this feature evolves during the decaying
phase of the flare towards the more usual value of 1 at lower
fluxes. This is the first time Compton dominance is observed in a HBL, 
suggesting a bimodality in the flaring mode of this kind of source: either
synchrotron or Compton dominated. Additionally, a cubic relation is
observed in the evolution of the flare between the VHE and X-ray
fluxes ($F_{\gamma} \propto F_{\rm{X}}^{3}$) which cannot be accounted for
by a single-zone SSC model. Both these properties indicate that the
entire SED behaviour could result from a superposition of two emission
zones: a steady component, responsible for the ``persistent''
emission, and a more compact one, with larger bulk motion, responsible for
the flaring activity via an external Compton channel in the interplay
between different parts of the jet.               
     
\subsubsection{H.E.S.S. and Fermi observations of the
  quiescent state in 2008} 

In August 2008, the H.E.S.S. and Fermi collaborations combined efforts in a MWL
campaign to sample the full Inverse-Compton SED of a TeV blazar,
observationally constraining for the first time the peak of the IC 
bump\cite{hessfermi}. During this campaign PKS 2155-304 was found to be
in a quiescent state in VHE. The fractional rms variability above 200
GeV was $\sim$ 23\% $\pm$ 3\%, and simultaneous X-ray RXTE data
also presented variations with flux doubling episodes in timescales of
days; the optical and the 0.2-300 GeV Fermi fluxes showed no
significant variations during the 11 nights of the campaign. 
The optical-to-TeV time-averaged SED of the source was
modelled by a single-zone SSC model which broadly fits the entire
profile (see Fig.~\ref{fig2}); the model parameters used for the fit
match those of an SSC description of a steady large jet component for
PKS 2155-304\cite{katarzynski}. A detailed study of the model reveals
that the most energetic electrons in the system are
responsible for the X-ray emission. These electrons
do not contribute to the GeV-to-TeV IC emission due to Klein-Nishina 
suppression and all gamma-ray emission can be roughly explained
as IC emission from a lower-energy particle population, responsible
for the optical sycnhrotron radiation. This SED justifies the
lack of correlation observed between the X- and gamma-ray fluxes --
different to that observed in the 2006 high-state -- but fails
to explain the lack of variability in the Fermi bands,
challenging the simple picture that both Sy and IC emissions are
due to the same particle population, and pointing again towards some kind of
multi-zone scenario.

\subsection{H.E.S.S. Observations of Radio Galaxies}

The detection of $\gamma$-rays from distant extragalactic objects is
possible due to effective acceleration and Doppler boosting of the 
radiation in relativistic outflows; in the abscence of this boosting,
such as in misaligned-jet sources, only extremely powerful, 
``isotropic'' sources are expected to be detectable at VHE
energies. Nearby radio-galaxies ($<$ 100 Mpc) such as M 87
($\phi_{\rm{l.o.s.}} \sim$ 20$^{\circ}$-40$^{\circ}$) and Cen A 
($\phi_{\rm{l.o.s.}}$ $>$ 15$^{\circ}$)
are part of this category, and have recently
been established as VHE emitters. The
importance of observing nearby sources at high-energies is that
they provide a unique opportunity to probe the origin of the
VHE emission with unprecedented spatial resolution. Given the
similarity among the systems, this might add to our understanding of gamma-ray
emission in more distant AGNs such as blazars.   

\begin{figure}[pt]
\centerline{\epsfig{file=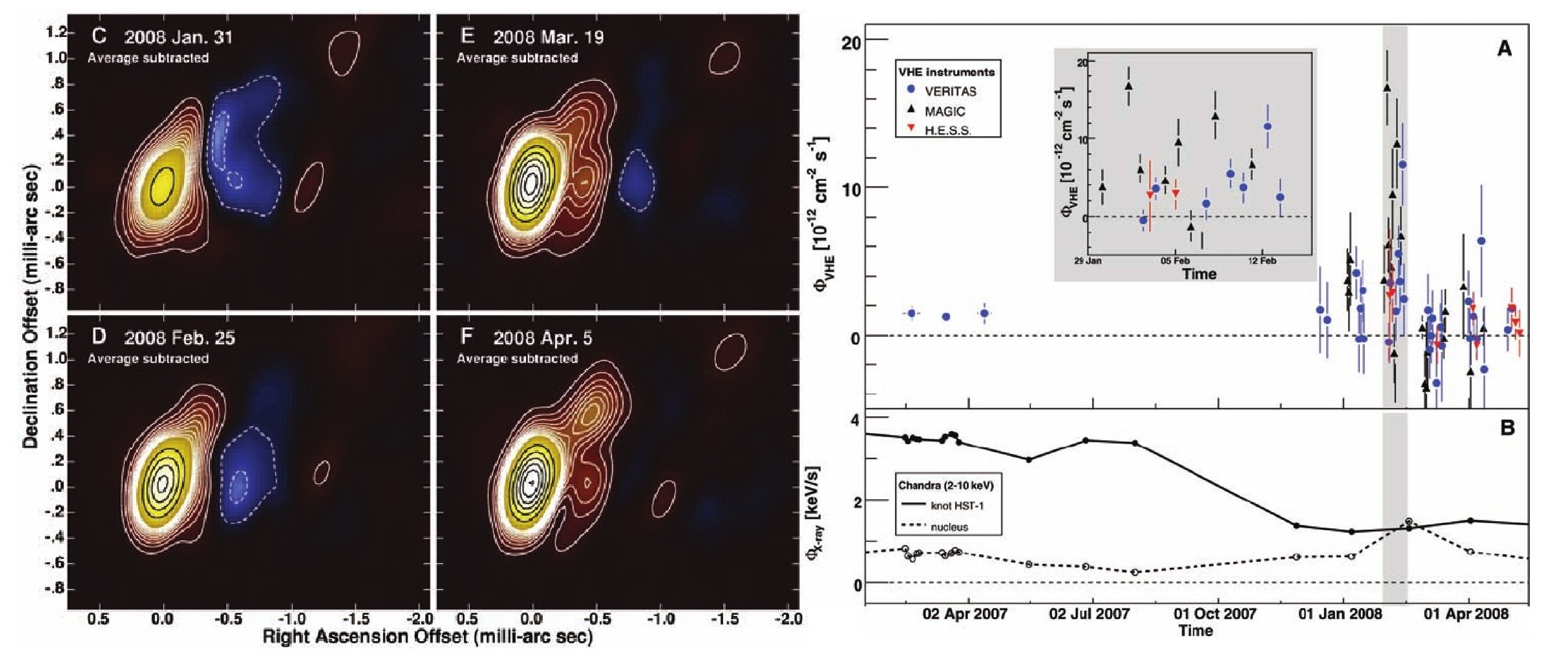, width=13cm}}
\vspace*{8pt}
\caption{Left: Sequence of VLBA images of M87 at 43 GHz during the
  period of the 2008 radio flare, showing a rise in the core flux
  density and the appearance of enhanced emission along the inner
  jet\protect\cite{walker}. Right: Nighly VHE (top) and Chandra (bottom) light curve of
  M87 for the 2008 joint campaign. The solid and dashed lines in the
  bottom panel represent the HST-1 and core X-ray flux, 
  respectively\protect\cite{hessfermi}. Reproduced by permission of the
  publisher.}
\label{fig3}
\end{figure}

\subsubsection{M87: a laboratory for jet and AGN physics}

M 87\cite{beilike} (16 Mpc; $M_{SMBH} \sim 3 \times 10^{9} M_{\odot}$)
was the first radio-galaxy observed at VHE energies\cite{hegra}, for which
H.E.S.S. detected a point-like $\gamma$-ray signal in positional
coincidence with the central regions of the object\cite{M87a}. Although
the source is close enough for the PSF of the telescopes to exclude the
outer lobes as the site of the emission, the 3 arcmin size upper limit for the
TeV source is consistent with several potential sites of particle
acceleration and $\gamma$-ray production, including the central AGN and
different regions of the inner kpc-scale jet\cite{biretta}. 
Observations of the source since 2003\cite{M87a}$^{,}$\cite{M87b} displayed 
flux variability on scales of years and days, constraining
the size of the $\gamma$-ray emission site to be very compact (with a
size U.L. $< 100R_{\rm{S}}$) and therefore probably located in
the vicinity of the central engine. 

The first VHE flare in 2005 was detected during an exceptional,
long-term high X-ray state of the knot HST-1, some 60 pc 
($\sim$ 1'') from the radio nucleus of the jet\cite{harris}, but
without unambiguous suggestion of a direct correlation existing
between the two events. Further investigations were conducted in a 
coordinated H.E.S.S., MAGIC and VERITAS campaign in 2008, in
which a stronger flux outburst with variability timescales of
days was clearly detected\cite{M87b}. Contemporaneous 43 GHz radio
VLBA\cite{walker} and X-ray Chandra\cite{harrisb} observations were
also conducted (see Fig.~\ref{fig2}). Unlike 2005, HST-1 was in a low
X-ray state whereas the core showed a flux 
increase peaking around the time of the maximum of the TeV activity. Enhanced
radio activity was also present in the core and the VLBI images
recorded the emergence of a new knot happening contemporaneously to the
flare at high energies. These more recent MWL results
suggest that the TeV emission in M 87 takes place in the immediate
surroundings of the SMBH, well within the jet collimation region\cite{junor}
(the radio core is $\sim 100 R_{S}$ downstream from the central
engine).

\subsubsection{H.E.S.S. detection of Centaurus A}

H.E.S.S. has recently reported the discovery of VHE $\gamma$-ray
emission from the radio galaxy Centaurus A\cite{israel} (3.8 Mpc), the
nearest active galaxy to Earth. The signal flux of $\sim$ 8 mCrab
($L = 2.6 \times 10^{39}$ erg s$^{-1}$) is associated with
a region including the radio core, kpc jets and the inner
radio lobes\cite{cenA}, the best fit position being well compatible
with the innermost regions: a scenario similar to the picture we have
for M 87.     

\section{Variable Galactic Sources}

LS 5039\cite{ls5039} and PSR B1259-63\cite{PSRB} are the only two variable
galactic TeV sources detected by H.E.S.S. unambiguously associated
with compact binary systems. The two systems differ in
the nature of the compact object: whereas PSR B1259-63 is known to be
a ``binary plerion'', it is debatable if in LS 5039 it is a spinning
neutron star or a black hole that is orbiting the massive 
companion\cite{dubus}. These two
possibilities leave the interpretation open as to whether the TeV emission
results from pulsar wind interaction as in PSR B1259-63 or if it is
the product of accretion, and we are therefore seeing the first
observational evidence of TeV emission from microquasars. In the
latter scenario, emission would come from relativistic jets of
particles emanating from the BH-accretion disc system in an analagous
way to what happens in active galaxies.\cite{mirabel} 

An interesting feature of the observations of these objects is that
their fundamental timescales are dictated by the orbital
period of the binary system (days-years), since the high magnetic
fields involved (up to $\sim$ 1 G) imply rapid electron cooling rates: one
expects therefore the emitted flux to more or less reflect the
(radically different along the orbit) ``instantaneous'' environment of
the system and the variability is modulated according to the orbital
configuration (i.e. apastron-periastron passages). In fact,
for both systems such modulation has been
observed.\cite{ls5mod}$^{,}$\cite{psrbmod}

Details of such variability depend on three factors: the radiation
and absorption mechanisms and the geometry of the system. In the case
of LS 5039, for which radio images suggest the presence of a
jet\cite{paredes}, the observed flux peaks at inferior conjuction,
when the compact object passes in front of the massive
companion. Flux variation is accompanied by spectral variability,
mostly probably as a result of $\gamma$-$\gamma$ pair production on
the radiation field of the massive companion. Nevertheless, a simple
picture where orbital modulation is solely the result of absorption
cannot explain the detailed features of the observation: the maximum 
spectral modulation is expected to happen at around 300 GeV
(temperature of the companion photon field $\approx$ 3 eV) whereas the
observed variation peaks at 5 TeV, with little signature of variability at the
predicted energy. Additionally, the flux at superior conjunction is
not zero, despite the expected high density of the
intervening photon field, and it remains to understand if this
residual flux is due to a particular geometrical configuration and/or
the result of direct observations of IC electromagnetic 
cascades.\cite{dubus}

Due to the high eccentricity of the orbit, the period of PSR
B1259-63 is relatively long ($\approx$ 3.4 years), and the system was
observed by H.E.S.S. during two periastron passages, in 2004\cite{PSRB}
and 2007\cite{psrbmod} (it is not detectable for orbital phases $\phi
\gg \pm 0.1$). The knowledge that it is a pulsar orbiting the
massive companion constrains the emission to be the result of pulsar-wind
interaction with the massive star's photon field. The variable
light-curve of the source shows peculiar behaviour, with two maxima
around a month before and after periastron passage. Mechanisms such as
IC scattering on the massive companion's photon field\cite{kirk} or
hadronic scenarios\cite{kawachi} (e.g. $pp
\rightarrow \pi^{0} \rightarrow \gamma \gamma$) have both been
suggested as possible origins for the TeV photons. But the most recent 
observations for the 2007 periastron passage, have detected an early 
($\phi > 0.05$) excess in the VHE flux which introduce new
difficulties for both approaches. The next periastron passage will
happen around October 2010.

\section{Prospects and Final Remarks}

H.E.S.S. has observed a number of different sources and source categories
whose emission is powered by particle acceleration in relativistic
outflows. Many of these objects are potential candidates for being
sources of Galactic and extragalactic cosmic-rays. Despite
improvements on our understanding of the origin of $\gamma$-ray
emission in the universe, the results presented here pose a similar
number of new and deeper questions on the nature of relativistic
sources. H.E.S.S.-II is expected to
greatly contribute to this work.\cite{hessII} The addition of a larger 
telescope to the centre of the array will decrease the energy
threshold of the system to $\sim$ 15 - 25 GeV and improve the
sensitivity of the instrument by a factor of 1.5 to 2 at
higher energies.         

\section*{Acknowledgments}

I thank my colleagues from the H.E.S.S. collaboration and the
Gamma-ray Astronomy Group at Durham for  
contributions to this review. The author
acknowledges a Ph.D. scholarship from the CAPES Foundation, Ministry
of Education of Brazil.



\end{document}